\documentclass[11pt,twoside]{article}

\usepackage{asp2006}
\usepackage{epsf}
\usepackage{lscape}
\usepackage{graphicx}

\usepackage{natbib}
\newcommand{\cdr}{\ensuremath{^{13}\mem{C}}}
\newcommand{\nezw}{\ensuremath{^{22}\mem{Ne}}}
\newcommand{\sprn}{\mbox{$s$ process}}
\newcommand{\spr}{\mbox{$s$-process}}
\newcommand{\kelv}{\ensuremath{\,\mathrm K}}

\newcommand{\abb}[1]{Fig.\,\ref{#1}}

\newcommand{\mem}[1]{\ensuremath{\mathrm{ #1}}}

\newcommand{\n}{\ensuremath{\mem{n}}}

\newcommand{\czw}{\ensuremath{^{12}\mem{C}}}

\newcommand{\nvi}{\ensuremath{^{14}\mem{N}}}
\newcommand{\msun}{\ensuremath{\, {\rm M}_\odot}}
\begin{document}
\title{Convective and non-convective mixing in AGB stars}   %%% Fill in title
\author{Falk Herwig\altaffilmark{1,2}, Bernd Freytag\altaffilmark{1,3},  Tyler Fuchs\altaffilmark{4}, James P. Hansen\altaffilmark{4}, Robert M. Hueckstaedt\altaffilmark{1}, David H. Porter\altaffilmark{4}, Francis X. Timmes\altaffilmark{1}, Paul R. Woodward\altaffilmark{4}}   %%% Fill in author names
\altaffiltext{1}{Los Alamos National Laboratory, Los Alamos, NM, USA}    %%% Fill in author affiliations
\altaffiltext{2}{Keele Astrophysics Group, School of Physical and Geographical Sciences, Keele University, UK}
\altaffiltext{3}{Centre de Recherche Astronomique de Lyon, Lyon, France}
\altaffiltext{4}{Laboratory for Computational Science \& Engineering, University of Minnesota, USA}

\begin{abstract} %%% Abstract to run on from here.

  We review the current state of modeling convective mixing in AGB
  stars. The focus is on results obtained through multi-dimensional
  hydrodynamic simulations of AGB convection, both in the envelope and
  the unstable He-shell. Using two different codes and a wide range of
  resolutions and modeling assumptions we find that mixing across
  convective boundaries is significant for He-shell flash
  convection. We present a preliminary quantitative analysis of this
  convectively induced extra mixing, based on a sub-set of our
  simulations. Other non-standard mixing will be discussed briefly.

\end{abstract}

\section{Introduction}   
Our understanding of the physics of mixing in stars
especially AGB stars \citep{iben:83b,herwig:04c}, is not in a
satisfying state. We do get many global properties right, but the list
of things that models can not accurately account for is getting longer
as the observations are becoming more detailed and numerous. The lack
of predictive models for mixing and consequent stellar properties and
nuclear yields is severely inhibiting the usefulness of this field for
helping to address some general questions in astronomy; for
example in the emerging field of near-field cosmology. With better
simulations we could use detailed abundance observations to precisely
characterize extra-galactic populations. Thus, there is a real need
for improving our understanding of convective and non-convective
mixing in AGB stars.

\begin{figure}
\begin{center}
\includegraphics[width=11cm]{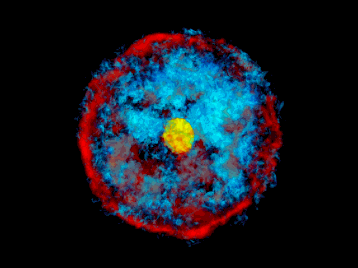}
\caption{Temperature fluctuations in AGB envelope convection simulations. See \citet{porter:00} and http://www.lcse.umn.edu/research/RedGiant for details.
\label{fig:ppm-env}}
\end{center}
\end{figure}
The most important physical process for mixing in stars is
convection. In 1D stellar evolution models we use the mixing-length
theory \citep{boehm-vitense:58} or some variant. The mixing-length
parameter determines the geometric scale of convective eddies and is
calibrated by using a solar model. However, we know for certain that
the mixing length parameter is not constant throughout all phases
and conditions in stellar evolution. Figure~5 in the important paper by
\citet{ludwig:99} shows the mixing-length parameter from their set of
2D radiation-hydrodynamic simulations of the outer convection layer in
stars similar to the sun. These simulations show that the
mixing-length parameter is very sensitive to the temperature (and thus
to the depth of the convection zone), and that just within the small
parameter range covered here, say within $\pm 1000\kelv$ the
mixing-length parameter changes by $\pm 15\%$. Such a variation would
change many important features in AGB simulations, most notably
dredge-up and the hot-bottom burning efficiency. This has been known since
first shown by \citet{boothroyd:88} and should be kept in mind when
comparing dredge-up efficiency-dependent model predictions and
observations. The hot-bottom burning dependence on the mixing-length
parameter has implications in particular for super-AGB stars and
massive low-metallicity AGB stars.

How can this situation be improved? \citet{canuto:91} and
collaborators have suggested the Full-spectrum turbulence convection
model as an alternative. This theory overcomes one important
simplification of MLT: convection does not only consist of one blob
size, but rather contains a spectrum of scales. Unfortunately, so far
it seems that no convincing validation case for AGB envelope
convection has been identified. Other uncertain input physics such us
mass loss, other sources of mixing (rotation and gravity waves), and
in some cases nuclear reaction rates have a similar, independent
effect on the same astrophysical observable (e.g. certain abundances,
like the \czw/\cdr\ ratio). In addition the variability of a
convection parameter studied by \citet[][see above]{ludwig:99} applies
here as well (see their Fig.~6).

A resolution to these problems can only come from simulations. Several
years ago \citet{porter:00} presented highly resolved simulations
with simplyfied input physics (\abb{fig:ppm-env}). One of their
important results was that convection settles into a dominant global
dipole mode, a finding that is now confirmed for fully convective
spheres in core-carbon burning WDs that are about to ignite a SN Ia
\citep{kuhlen:06}. Another finding was that convection itself without
any other physical mechanism leads to pulsations. These may match
observed pulsations of AGB stars, but a more detailed comparison needs
to be done. Other AGB envelope models in 3D were developed by
Freytag \citep[see][and http://www.astro.uu.se/$\sim$bf for more
details]{freytag:03,hoefner:05}. These models feature a more realistic
outer boundary condition but are not as highly resolved as the Porter
\& Woodward simulations.

\begin{figure}
\begin{center}
\includegraphics[width=12.5cm]{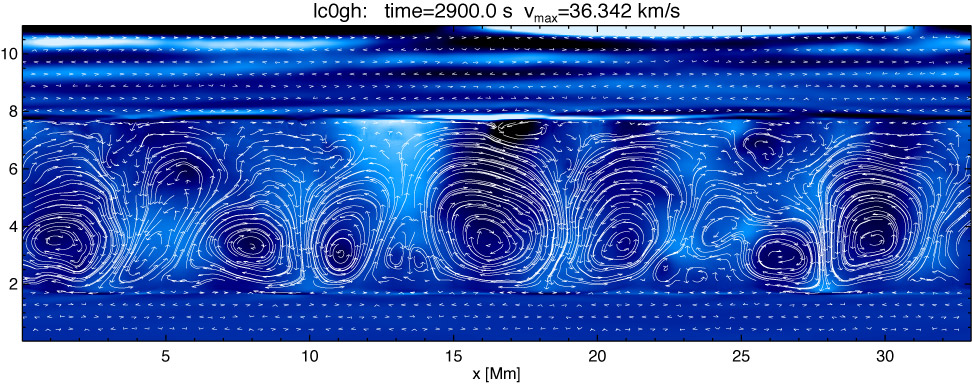}
\caption{Pressure fluctuations and pseudo-streamlines for 2D He-shell flash RAGE simulation snapshot at $t=2900\mem{s}$.
\label{fig:rage-pfluct}}
\end{center}
\end{figure}
Porter \& Woodward derived a mixing-length parameter equivalent to the
velocity field from the simulations. They find $\alpha_\mem{MLT} \sim
2.6$. These simulations, both by Freytag and by Porter \& Woodward,
did not resolve the inner parts of the envelope convection, which is
so crucial for nucleosynthesis. We anticipate that significant
improvements of the treatment of convection properties in stellar
evolution modeling can be made through hydrodynamic models with
available and coming resources.
\begin{figure}
\begin{center}
\includegraphics[width=13cm]{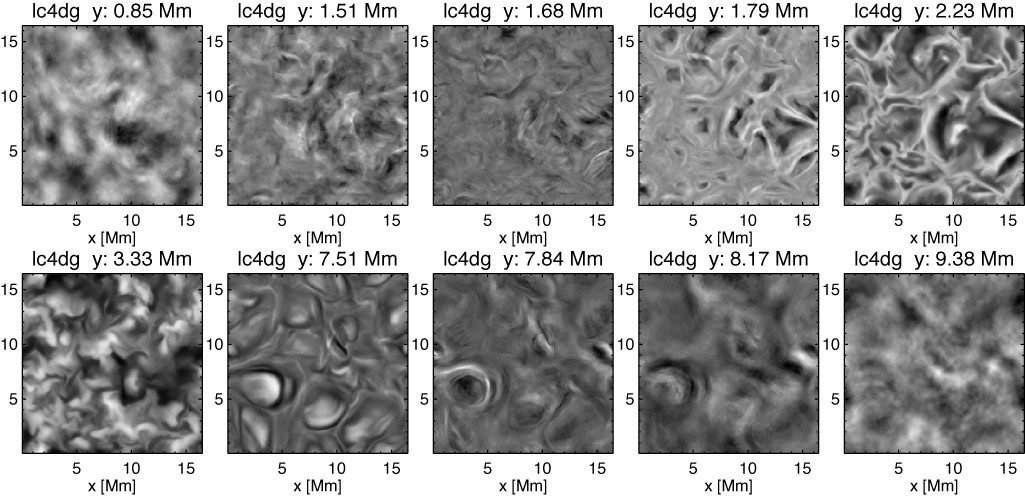}
\caption{Horizontal planes of vertical velocity at different vertical
locations of 3D RAGE run lc4dg ($300^2 \times 200$) at
$t=3330\mem{s}$. Panels for the vertical position $y=1.79, 2.23, 3.33,
7.51 \mem{\ and\ } 7.84 \mem{Mm}$ are inside the convectively unstable zone.
\label{fig:rage-lc4dg-v3}}
\end{center}
\end{figure}

\section{Multi-D hydrodynamic simulations of He-shell flash convection}
\subsection*{Motivation, codes and setup}
In order to understand better nucleosyntheis in AGB stars, we have
performed multi-dimensional He-shell flash convection
simulations. This work has several goals. First of all, the
hydrodynamics of He-shell flash convection has never been simulated
before. So, we want to investigate the hydrodynamical properties and
the topology of He-shell flash convection. For example, how well does
MLT describe the vertical velocity profile of the convection zone?
What are the dominating scales? What is the dependence on resolution
and on the nuclear energy generation?  Obtaining a resolved velocity
distribution from hydrodynamic simulations can give a new framework to
study short-lived, temperature-dependent s-process branchings fueled
by the \nezw\ neutron source. And, most importantly, we want to study
convection induced mixing across the convective boundaries. Several 1D
stellar evolution studies using a parameterized recipe for convective
overshooting \citep[e.g.][]{herwig:99a,lugaro:02a} have shown that
overshooting at the bottom of the He-shell flash convection zone
strongly affects the pulse strength, and thus the dredge-up
efficiency, as well as the intershell abundance and the temperature
for the \nezw\ s-process. Some of the predictions of stellar evolution
models with convective extra mixing at the bottom of the He-shell
flash convection zone have been confirmed by observations of
H-deficient post-AGB stars \citep{werner:06}.

We have performed simulations using two codes. One set of simulations
\citep[see][for initial results]{herwig:06a} was done with the RAGE
code, an explicit, compressible, Eulerian grid code. The other code (scPPM) is
a PPM (Eularian, compressible grid as well) code for stellar convection that
solves for perturbations to a base state defined in the problem
setup. This later approach takes advantage of the fact that buoyancy
driving the convection in these stellar interior environments is very
small and that Mach numbers of the flow are small ($\mem{Ma} <
0.03$). We will give more details on the scPPM code in a forthcoming
paper.

\begin{figure}
\begin{center}
\includegraphics[width=6.cm]{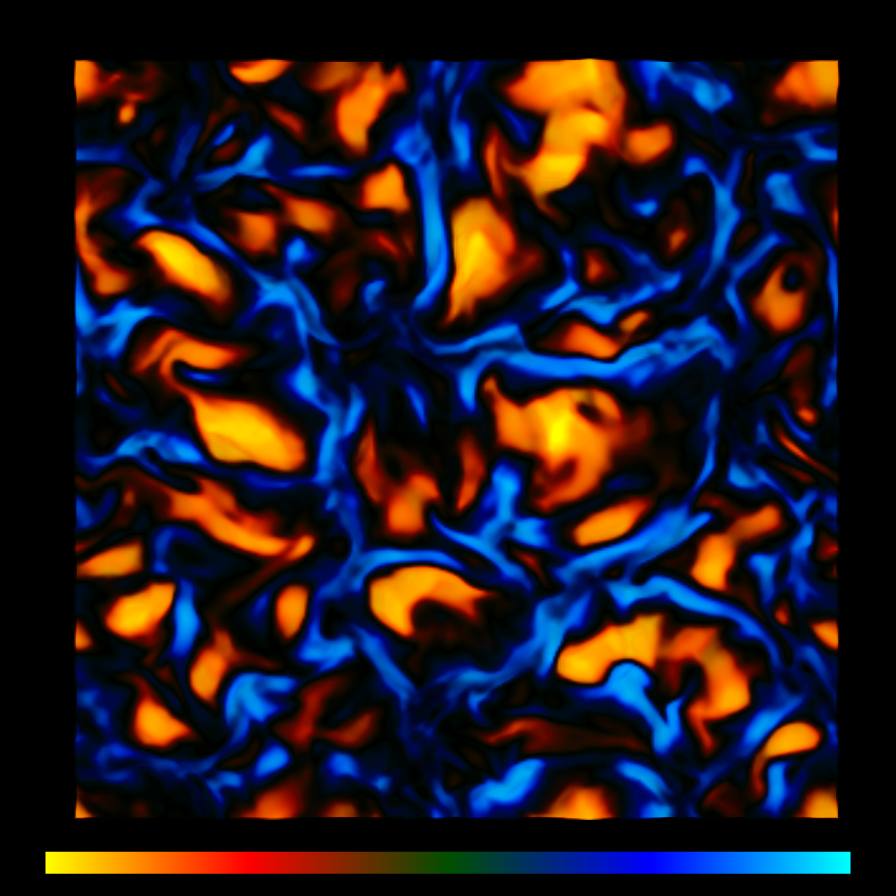}
\includegraphics[width=6.cm]{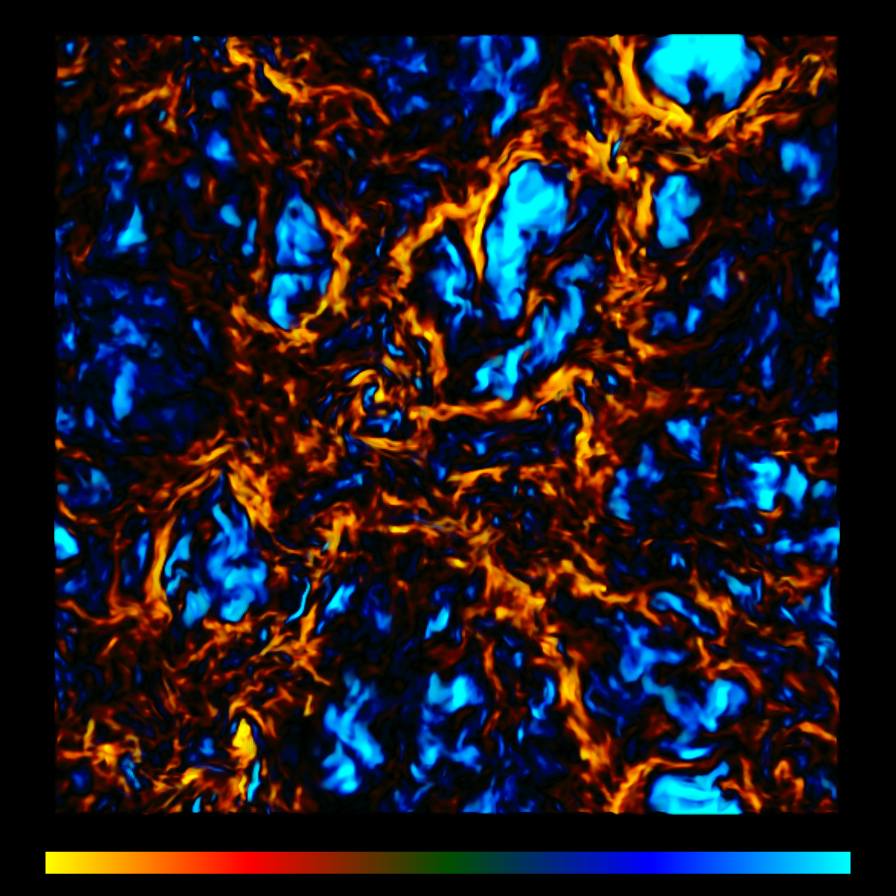}
\caption{Volume rendering of the vertical velocity in a horizontal
  slab with an approximate thickness of $\sim 1200\mem{km}$
  (corresponding to $20\%$ of the convectively unstable layer)
  including the top of the convection zone for the RAGE $300^2\times 200$
  (left) and the scPPM $512^2 \times 256$ (right) simulation at
  $t=1990\mem{s}$. The view is from top. Blue are positive (upward)
  and yellow are negative (downward) velocities.
  \label{fig:convtop_thin}}
\end{center}
\end{figure}
For the initial setup we construct a piecewise polytropic
stratification with gravity that resembles closely the actual
conditions in a detailed $2\msun, Z=0.01$ thermal pulse stellar
evolution model just before the He-shell flash peak luminosity. The
driving of the convection results from a constant volume heating at
the bottom of the convectively unstable region that injects the same
amount of energy as the corresponding stellar model.

The stratification comprises approximately 11 pressure scale heights,
half of which represent the convectively unstable region. This gives
enough simulation space for stable layers both above and below the
convection zone to avoid artifacts in the convection boundary
simulation from simulation box boundary effects.

\subsection*{Morphology and gravity waves}
In \abb{fig:rage-pfluct} we show a snapshot from a 2D RAGE run on a
$1200\times400$ grid. The same heating rate as in the stellar evolution model
was applied. At the snapshot time the simulation has lasted for
approximately 5 convective turn-over times and has thus reached
convective steady-state.  The initially imposed convective boundaries
are well maintained, sharp and clearly visible in this snapshot. The
convective flow is dominated by three to four large systems, which
vertically span the entire unstable layer and are centered in the
lower half of the unstable region. In the stable layers above and
below the convection zone oscillations due to internal gravity waves
can be seen.  Through a systematic resolution study we find that the total
number of large convective systems does not change over a wide range
of grid sizes, down to a $300\times100$ grid.

Gravity waves can be identified in the top and bottom stable
layers. Movies of these simulations show that the g-modes in the
stable layer above the convection zone are excited even before
the convective motions have reached the top boundary of the unstable
layer. In contrast to shallow surface convection, for example in
A-type stars \citep{freytag:96}, coherent convective systems do not
cross the convective boundaries in a fashion visible in this
representation. This is not surprising, because the relative stability
of convective boundaries in the stellar interior is much larger than
in stellar near-surface convection.

We have performed 3D simulations with both the RAGE and scPPM
code. The vertical velocities in one snapshot of the 3D RAGE run are
shown in \abb{fig:rage-lc4dg-v3}.  In the He-shell flash convection
zone, convection originates through heating at the bottom, contrary to
surface or envelope convection which is driven by cooling from the
surface. It is interesting to note that here we see near the bottom of
the convection zone (panel 2.23Mm) structures that resemble the
inverse of solar granulation: cool granules with hot intergranular
lanes.

\begin{figure}
\begin{center}
\includegraphics[width=12.5cm]{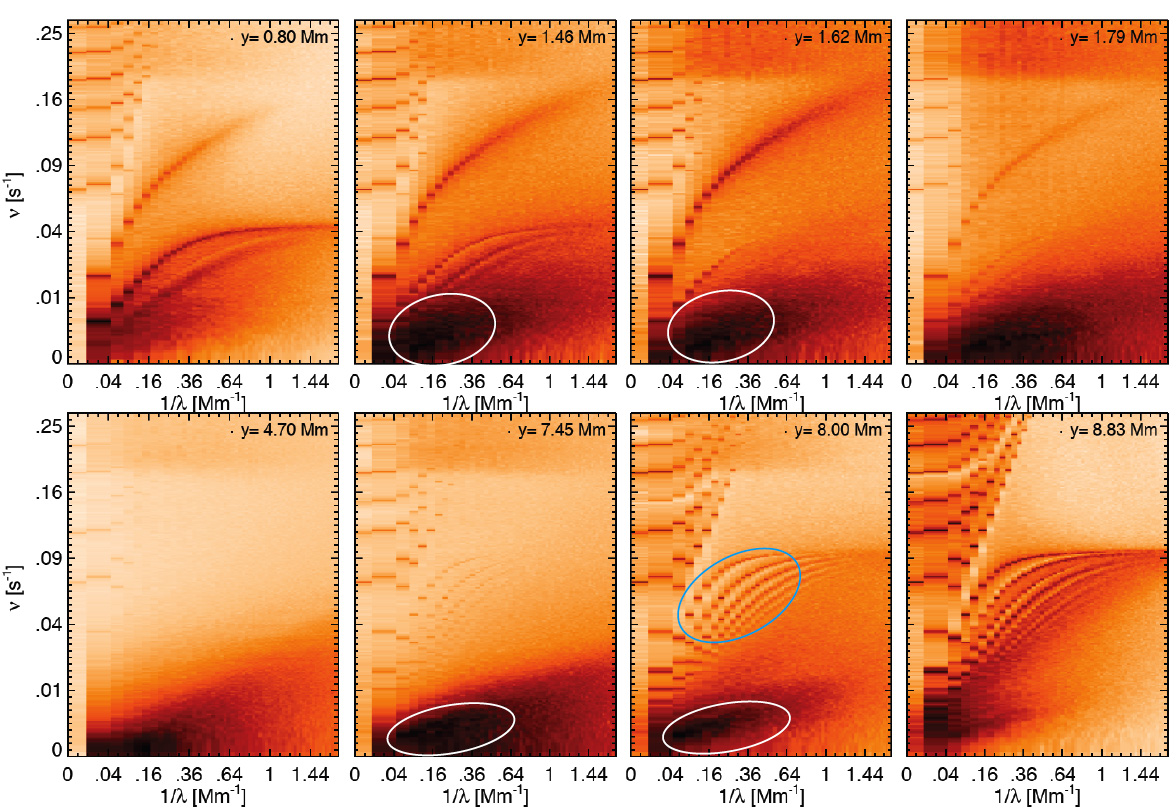}
\caption{k-$\omega$-diagrams for horizontal planes of the 3D RAGE
  run. Panels $y=1.62, 1.79, 4.70 \mem{\ and\ } 7.45\mem{Mm}$ are in
  the unstable layer. The signature of convection is a prominent blob
  at low wave number and frequency marked with white ellipses. g-modes
  can be identified best in the panels representing stable layers, and
  are marked with blue in the panel for $y=8.00\mem{Mm}$. p-modes are
  visible in the far left columns of the panels.
  \label{fig:k-omega}}
\end{center}
\end{figure}
Even outside the convection zone dynamic fluid motions exist, and
their patterns again depend on the distance to the convective
boundary. Panel 1.51Mm shows a very granular appearance with a lot of
detail on small scales whereas both the panels above and below show
fluctuations on larger scales. The layers most distant from the
convection zone, both above and below, show a rather diffuse pattern
of large-scale fluctuations.

\begin{figure}
\begin{center}
\includegraphics[width=10cm]{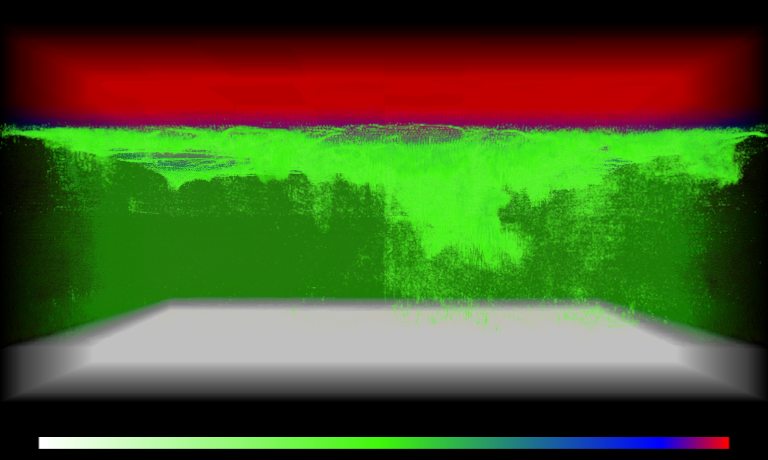}
\caption{Concentration of fluid that is initially present only
in the stable upper layer, at $t=1510\mem{s}$ in the $512^2 \times 256$ scPPM
convection simulation. Color coded is the log of the concentration,
from $0$ (red) to $-4$ (white). Rendering opacity is sharply peaked
around $-2$ (green).
\label{fig:cr3Fal0151}}
\end{center}
\end{figure}
Although the coherent convective systems do not cross the convective
boundary on a scale that is any significant fraction, of the
horizontal scale, this does not mean that there is no oscillation
exitation and mixing across the boundary. Looking at the vertical
velocity reveals how motions and oscillations in the stable and
unstable layer do in fact correlate, and therefore
communicate. Features in one slice have corresponding features in
neighboring slices above and below inside the convection zone, even if
the dominating scales are different in different layers of the
convection zone. This is consistent with the observation in the 2D
snapshots that convective system traverse the entire convection zone
vertically. However, patterns in the vertical velocity images also
show correlations between the boundary planes just inside the
convection zone and the neighboring planes several hundred km out in
the stable layer. Thus, while convective systems do not cross the
convective boundaries, they do imprint their signature on the
oscillation properties of the stable layers.

But how do properties like the vertical velocities depend on numerical
resolution and numerical scheme? To address this question we compare
in \abb{fig:convtop_thin} a snapshot of the velocity field in the
upper layers of the convection zone from the RAGE and scPPM
simulations. While the scPPM simulation shows much more small scale
structure, there are similarities in the overall pattern. Downflows are
ordered in lanes surounding larger areas with upwelling flows. The
more highly resolved grid of the scPPM simulation is only partly
responsible for the difference. The main reason for the difference in
scale distribution is the higher order of the PPM numerical scheme
compared the modified Van Leer method implemented in the RAGE version
used here.

The correspondence between different layers, including nearby
convective and non-convective layers, can be further analysed in a
series of k-$\omega$ diagrams of several vertical planes in the 3D RAGE run
(\abb{fig:k-omega}). These diagrams show the distribution of power as a
function of wavelength and frequency. In these diagrams the signature
of convective motions shows up as an unstructured blob in the lower left
part of the diagram. It is most prominent in the panels showing planes
in the unstable region. However, even in the panels representing the
stable layer a significant convective signature is
evident. In addition, the stable layers show the characteristic
signature of gravity waves, and all panels show the signature of p-modes.

We conclude that our simulations show that convection
does influence the fluid flows in the neighboring stable layers, both
through exciting g-mode oscillations and by forcing motions with
convective wavelengths and frequencies. We now need to look at how
these correspondences between fluid flows on both sides of the
convective boundary translate into actual mixing.

\subsection*{Mixing for He-shell flash convection}
\begin{figure}
\begin{center}
\includegraphics[width=8cm]{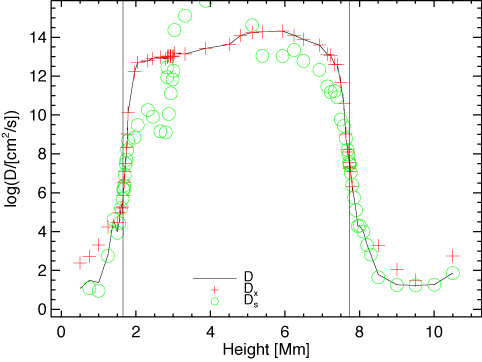}
\caption{Diffusion coefficient as a function of vertical position derived with the tracer particle method
  from the velocity fields of a 2D $600 \times 200$ run.  Green
  circles and red crosses are diffusion coefficients using spread
  evolutions of entropy and vertical coordinate respectively. The
  solid black line is the composite final result (see text).
\label{fig:pathdiff_lc0gg_Ddiff}}
\end{center}
\end{figure}
The scPPM simulation has been performed with multiple fluids
representing the different abundance compositions in the three layers
of our sandwich setup. In \abb{fig:cr3Fal0151} we show the result of
material from the top stable layer above the convection zone being
entrained into the convection zone. Conceptually, such a situation may
occur in He-shell flashes of extremely metal-poor stars when the
convection zone reaches into the H-rich envelope, or in the very-late
thermal pulse flash associated with the born-again evolution. In this
simplified simulation, the effect of $\mu$-gradients has been neglected
as all fluids have the same molecular weight. Therefore the
entrainment seen in this case is probably an upper limit. However, we
do see that the distribution of the fluid from the upper stable layer
in the convection zone is rather isotropic. If the fluid from above
would react with the material in the convection zone we would likely
be in a distributed burning regime, rather than in a flame propagation
situation. Using the multi-fluid scPPM code we will in the future be
able to quantify mixing across the convective boundary. In this way we
can take full advantage of the high-order advection scheme implemented
here \citep{woodward:06}.

With the RAGE simulations we performed a mixing analysis based on the
tracer particle technique of \citet{freytag:96}. We determined 
diffusion coefficients at different horizontal positions, in particular
near the convective boundaries. An example for this approach (full
details will be given in Freytag etal. 2006, in prep) is shown
\abb{fig:pathdiff_lc0gg_Ddiff}.  The final solution of the diffusion
coefficient as a function of vertical position is a composite of two
approaches. Diffusion coefficients derived from the spread evolution
of the entropy are more suitable for the stable layer. The evolution
of the vertical coordinate is more appropriate in the convection
zone. In the transition layer, both approaches give the same
result. The fall-off of the diffusion coefficient has been fitted
with an exponential. This approach has before been used by
\citet{freytag:96} and applied to stellar evolution calculations,
first by \citet{herwig:97}. Within this framework, the convection
induced mixing across the bottom convective boundary can be described
by a succession of two exponential decay laws of the form $D =
D_0\exp{-2z/fH_\mem{p}}$ where $z$ is the distance from the location
at which a baseline $D_0$ has been determined, and $H_\mem{p}$ is the
pressure scale height \citep{freytag:96,herwig:99a}. The first
decay starts already somewhat inside the convection zone with
$f_\mem{bot,1}=0.01$, while just outside the convection zone (when D
has fallen to $\sim10^5\mem{cm^2/s}$) the decay of mixing efficiency
flattens and can be represented by $f_\mem{bot,2}=0.14$. At the top
boundary our initial analysis gives a single exponential decay with
$f_\mem{top}=0.10$.

The f-value determination varies as a function of resolution within a
factor of a few. However, we could not find any apparent trend,
indicating that resolution may be a limiting aspect within the
chosen range of simulations. Also we did not find a systematic
difference between simulations with the realistic heating rate
compared to those with 30 times larger heating rate. This indicates
that within a range of driving energies, mixing properties will depend
predominantly on the background stratification. This means that the
f-value for mixing across the convective boundary remains the same for
a given convective boundary until the conditions change drastically.

From the few 3D results we have, we
find that the 2D/3D difference is of the same order as the dependence
on resolution. Overall the 3D runs confirm the results we found in 2D.  

Initial stellar evolution test calculations over a couple of thermal
pulses implementing these hydrodynamic convective mixing results
indicate that we recover the increased O and C intershell abundance
that we found in earlier calculations using the f-overshoot for this
convection zone. The simulated C and O abundances in the intershell
are consistent with observations of H-deficient PG1159 and [WC]-CSPN
stars \citep{werner:06}.

\section{Non-convective mixing}
Rotation has an important effect on the production of elements in AGB
stars but is unfortunately not yet studied very well. Models of
rotating AGB stars \citep{langer:99} show that with the present
physics model rotation induced mixing at the convective boundary is
insufficient for the formation of the \cdr-pocket for the \sprn. Even
worse, \citet{herwig:02a} showed that rotationally induced mixing may
be even harmful to the \spr\ nucleosynthsis. In models in which a
\cdr\ pocket was introduced using the convective overshooting
paradigm, rotation induced mixing of \nvi\ into the \cdr-pocket during
the interpulse phase. The neutron exposure is well below $\tau
=0.1\mem{mbarn^{-1}}$ which is far too low to reproduce well
established s-process observables. \nvi\ acts as a neutron poison due
to its large $(\n,p)$ cross section. Overall, there are indications
that angular momentum transport from the core to the envelope is
underestimated in current models of rotating AGB stars.  Therefore
angular velocity gradients at the core-envelope interface, and
consequently rotation induced mixing at this interface are
overestimated. The \cdr-pocket is located exactly at this
interface. Another indication for missing angular momentum transport
is that core rotation rates of rotating AGB models are too large compared
to observed rotation rates from white dwarf pulsations
\citep{kawaler:03}.

Magnetic fields could add angular momentum transport and reduce
angular velocity gradients that cause shear mixing. However, there is
another likely possibility, in particular in light of what we see in
our hydrodynamic simulations. Internal gravity waves could have a
similar desired effect of angular momentum transport (Talon \&
Charbonnel, this volume). We see them plentifully in the hydrodynamic
simulations, and their effect on mixing in AGB stars has not yet been
studied in sufficient detail.  The only work on gravity waves in
AGB stars besides the new results by Talon \& Charbonnel is that by
\citet{denissenkov:02} who show that mixing induced by these waves may
provide the conditions needed for the formation of a
\cdr-pocket. However, more detailed studies in this area are needed.

\section{Conclusions}
Hydrodynamic simulations of convection in AGB stars which give
meaningful insight for stellar evolution models, both in the envelope
and in the intershell are feasible and offer an exciting tool to
study mixing. We can obtain a detailed picture of flow structures both
within the unstable zone as well as in the neighboring layers. We will
study in greater detail the variation of the averaged mixing
efficiency for AGB envelopes that determines important evolutionary
properties, such as the dredge-up efficiency as well as the hot-bottom
burning efficiency.  Our initial results on He-shell flash convection
allow a first quantitative glimpse at mixing at and across the
convective boundaries. The simulations emphasize the need to study the
role of gravity waves in much greater detail. The issue of
rotationally induced mixing is much more difficult to approach with
multi-dimensional simultions, mainly because the flow velocities are
significantly smaller than in convection. This is especially
unfortunate as rotating models currently do not reproduce observables of
AGB stars although these stars obviously rotate, at least their cores.

\acknowledgements 
This work was carried out in part under the auspices of the
        National Nuclear Security Administration of the
        U.S. Department of Energy at Los Alamos National
        Laboratory under Contract No.\ DE-AC52-06NA25396.

%\bibliography{astro}

\end{document}